# Using real-time cluster configurations of streaming asynchronous features as online state descriptors in financial markets

D. HENDRICKS*

School of Computer Science and Applied Mathematics, University of the Witwatersrand
Johannesburg, WITS 2050, South Africa

(*v1.1 released April 2017*)

We present a scheme for online, unsupervised state discovery and detection from streaming, multi-featured, asynchronous data in high-frequency financial markets. Online feature correlations are computed using an unbiased, lossless Fourier estimator. A high-speed maximum likelihood clustering algorithm is then used to find the feature cluster configuration which best explains the structure in the correlation matrix. We conjecture that this feature configuration is a candidate descriptor for the temporal state of the system. Using a simple cluster configuration similarity metric, we are able to enumerate the state space based on prevailing feature configurations. The proposed state representation removes the need for human-driven data pre-processing for state attribute specification, allowing a learning agent to find structure in streaming data, discern changes in the system, enumerate its perceived state space and learn suitable action-selection policies.

*Keywords*: state space discovery; online learning; asynchronous data; financial markets; market microstructure

## 1. Introduction

Machine learning has become ubiquitous in high-frequency financial markets, as technological advances enable low-latency automated algorithms to replace functions traditionally performed by human traders, portfolio managers, risk managers and regulators. This is particularly true for trading algorithms, where reinforcement learning algorithms have recently been considered as dynamic alternatives to traditional stochastic control techniques (such as those found in Bertsimas and Lo (1998), Almgren and Chriss (2000), Forsyth (2011), Frei and Westray (2015), Cartea et al. (2015)) for mapping optimal trajectories through the system. Nevmyvaka (2004), Nevmyvaka et al. (2006) were among the first authors to consider a reinforcement learning agent for optimal limit order placement for a liquidation program. They used a discrete-state, discrete-action *Q-learning* agent which converged to a policy for the optimal price at which to post the remaining inventory in the market, based on the time remaining in the liquidation program, remaining inventory to trade and domain-informed public state attributes, such as prevailing spreads, price levels and volumes. Hendricks and Wilcox (2014) considered a similar problem, demonstrating that a reinforcement learning agent can be used to adapt a trading strategy with respect to prevailing spread and volume dynamics, executing a sequence of optimised market orders. These authors demonstrate a significant positive improvement on cost of trading compared to state-of-the-art techniques, motivating reinforcement learning as a suitable framework for online learning agents in financial markets. Both studies, however, use a subjective set of attributes for state representation in the learning algorithm. While the choice is informed by domain knowledge and may be suitable at the operating scale of a human trader, we conjecture that a more objective representation may yield better trading policies for agents operating at machine scale.

---

*Current affiliation: Machine Learning Research Group, Oxford-Man Institute of Quantitative Finance, University of Oxford. Email: dieter.hendricks@eng.ox.ac.uk

It is well known that the performance of certain classes of machine learning algorithms is strongly dependent on the choice of data representation, or features, upon which they are applied (Hinton (2007), Lee et al. (2009), Bengio et al.). This is likely due to certain forms of representation masking exploitable characteristics explaining variations in the data, or at least burying them in layers which cannot be detected by the learning algorithm. As such, significant effort can be spent on data pre-processing, using domain knowledge to inform appropriate representations for effective machine learning. While such human intervention can be useful to guide learning agents in new domains, it does restrict the agent's discoverable policies to those which mimic policies acceptable to an intuitive *human* agent in the domain. Bengio et al. state that an artificial intelligence should fundamentally understand the world around us, and thus be able to identify and disentangle explanatory features from low-level sensory data without human intervention. In this way, a machine learning agent can provide more general, and sometimes complementary optimal policies to those expected by human agents, thereby cultivating its own distinct *machine intelligence*.

This goal has been recognised by the machine learning community, with a recent surge in scientific activity concerning unsupervised feature learning (or deep learning), seeking the discovery of useful representations which result in more meaningful classifiers and predictors in various domains (see Dahl et al. (2012), Hinton et al. (2012), Boulanger-Lewandowski et al. (2012), Ciresan et al. (2012), Glorot et al. (2011), Krizhevsky et al. (2012) for some state-of-the-art examples). At a recent NIPS workshop, Mnih et al. presented the first deep learning model to successfully learn control policies from high dimensional sensory data using reinforcement learning Mnih et al. (2013). The agent was able to learn to play several Atari2600 games, using a convolutional neural network (CNN) trained using a *Q-learning* algorithm, with only raw pixels as the input. While this is a somewhat different domain to the optimal trade execution problem, it does present certain analogues consistent with our goal: using low-level sensory data (*pixels* here, vs *streaming tick data* for our problem), the CNN is able to abstract useful representations from the raw data and train a *Q-learning* agent to achieve some goal. While it would seem appropriate to apply this technique to our problem, the computational burden of the CNN may be too onerous for our goal of an online near-real-time algorithm using modest hardware. Even recent work of Ciresan et al. (2011) on state-of-the-art, computationally efficient, GPU-optimised CNNs yield computation times of the order of minutes for relatively simple problems.

We are thus tasked with developing a form of state representation which can be constructed directly from raw asynchronous tick data, is able to capture salient features of the limit order book, is computationally efficient for near-real-time use (of the order of seconds) and can be successfully combined with *Q-learning* for optimal trade execution policies. In the following sections, we describe our approach which is able to construct a rich state representation in a computation time of the order of seconds, using relatively modest hardware, enabling near-real-time state detection for online learning.

## 2. Cluster configurations as temporal state descriptors

In the previous studies considering state representations for high-frequency financial markets, the following pre-processed attributes were used as candidate descriptors: *bid/ask spread, quote volumes, quote volume imbalance, trade price levels* and *traded volumes* (Nevmyvaka et al. (2006), Hendricks and Wilcox (2014)). These were informed by common notions and intuition from human traders regarding the typical drivers of trade execution cost when interacting with financial markets. Consider a trader who is only able to execute *market orders* to satisfy an *arrival price objective* in a *limit order book* market. The *limit order book* is a schedule of quoted prices and volumes at which market participants are willing to transact, where *ask quotes* refer to *sell orders* and *bid quotes* refer to *buy orders*, with *bid quote prices* strictly less than *ask quote prices*. Participants who place limit orders effectively achieve can achieve a favourable price if they are willing to wait for the market to move in the direction of their limit price. There is thus uncertainty in the time at which the transaction will take place, if at all. Alternatively, a trader may place a *market order*, whereby they guarantee execution by matching against prevailing limit orders in the system, i.e. a *buy market order* will match against commensurate *limit ask quotes*, with the trade price calculated as the volume-weighted price of the matched ask quotes. The cost of this timing guarantee is thus paying a higher (lower) price for a buy (sell) order, where the extent of this cost is governed by the prevailing *quote depth* in the limit order book. The difference between the highest bid quote price and the lowest ask quote price is referred to as the *spread*.

Given the objective of minimising trading cost with respect to an arrival price benchmark, and the constraint of only executing market orders, a rational trader will attempt to plan his execution trajectory such that he *crosses the spread* infrequently, when there is sufficient quote volume at the *top of the limit order book* and the *spread* is narrow. This would minimise the cost paid for guaranteed execution. Thus, *spread* and *quote volume* were natural candidates for public state attributes if we wish our RL agent to learn this behaviour. In particular, when *spreads* are *narrow (wide)* and *volumes* are *high (low)*, we expect the RL agent to trade *more (less) aggressively* to minimise the trading program's overall execution cost.

While informed by domain knowledge and consistent with the data-preprocessing paradigm described in Bengio et al., these choices are somewhat subjective and are only capable of a partial representation of the true state space. Indeed, even the enumeration of all possible *spread* and *volume* configurations at the finest resolution is unlikely to be able to capture the endogenous and exogenous dynamics of the financial system, especially given recent arguments for multi-level causation and scale-specific behaviour (Wilcox and Gebbie (2014), Hendricks et al. (2016a)). While we can increase the complexity of the state space representation by increasing the number of attributes, the *curse of dimensionality* soon prevents computational tractability for an online algorithm, at least in the *Q-learning* setting we consider.

We propose an alternative notion to characterise the state at each decision point, effectively reducing the set of public attributes to a single metric, while preserving information from *all* measurable aspects of the system.

One can think of a particular realisation of state attributes as a cluster configuration of observable features for a stock. Consider the case of the model used by Hendricks and Wilcox (2014), where *spread* and *quote volume* were used as public attributes. These are derived from the following low-level features of the limit order book: *Level-1 Bid Price, Level-1 Bid Volume, Level-1 Ask Price, Level-1 Ask Volume*. Figure 1 illustrates how a cluster configuration of these low-level features has an analogous interpretation to the *low/high spread, low/high volume* regimes described in Hendricks and Wilcox (2014).

In time period $t_1$, we see *Level-1 Ask Volume, Level-1 Bid Volume* and *Level-1 Bid Price* are all correlated and increasing, thus being ascribed to the same cluster. *Level-1 Ask Price* is decreasing and is ascribed to another cluster. In particular, we notice that *Level-1 Bid Price* is *increasing* and *Level-1 Ask Price* is *decreasing*, which is consistent with a *narrowing spread* regime. Since we are considering market orders for a *BUY* trading program, we note that the narrow spread is accompanied by a larger *Level-1 Ask Volume*, which presents favourable conditions for an increase in trading activity. Thus the *low spread, high quote volume* regime considered in the *SSRQ* model has an analogous feature cluster configuration interpretation.

As a further example, consider the cluster configuration in time period $t_2$. Here, *Level-1 Ask Price* is increasing, while *Level-1 Bid Price* and *Level-1 Ask Volume* are both decreasing, resulting in a *high spread, low quote volume* regime, consistent with a decrease in trading activity.

This simple illustration demonstrates that the cluster configurations of low-level sensory features in high-frequency financial markets may have an analogous interpretation to the trader-intuition-derived regimes usually specified. Furthermore, by allowing the clustering algorithm to be exposed to streaming data from *all* measurable features, unique and persistent cluster configurations may yield meaningful state representations for machine learning classifiers and predictors, beyond those which may have been expected and proposed by human traders. In addition, an appropriate cluster configuration similarity metric can be used to identify temporal states which are characterised by the same feature cluster configurations. If certain configurations persist throughout the trading day, we then have a reduced number of states for which to solve our optimal trading policy. Thus, we will investigate using cluster configurations to describe temporal regimes in a reinforcement learning framework, utilising a high-speed cluster detection algorithm appropriate for online learning.

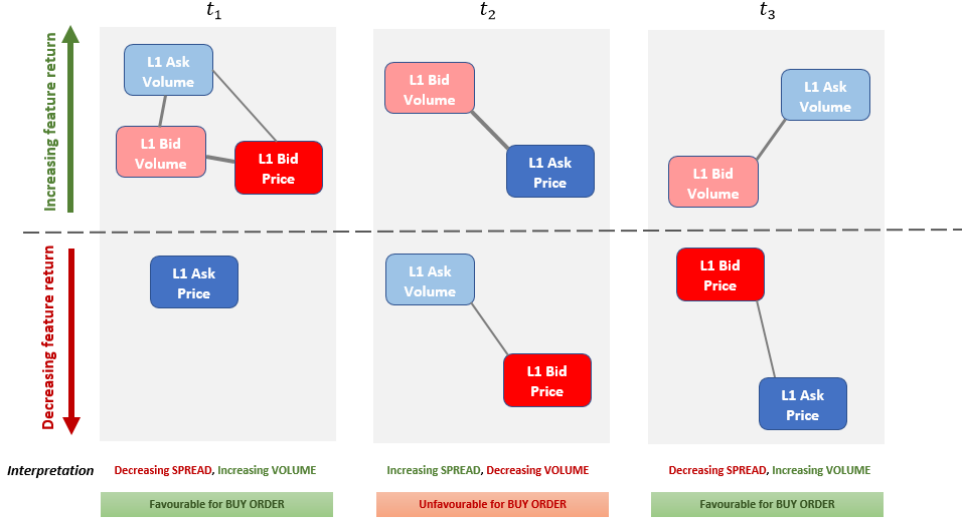

Figure 1. Illustrating how identified feature configurations may have an analogous interpretation, in terms of human-specified pre-processed features.

## 3. Correlation estimation from streaming asynchronous data

A key input in the clustering algorithm used in this analysis is an object *correlation matrix*. Classical estimators for co-volatility, and hence correlation, typically rely on evenly-spaced, synchronous observations for computation. In the case of high-frequency data in financial markets, the asynchronous arrival of the price time series would require the use of pre-processing or interpolation techniques before classical techniques can be applied, potentially introducing bias into the results (see Griffin and Oomen (2011) for a survey of candidate estimators). Malliavin and Mancino (2002, 2009) introduced a non-parametric co-volatility estimator based on Fourier series analysis, principally relying on the integration of a time series, rather than its differentiation. Their method removes the need for any artificially-imposed synchronicity, providing a measure of co-volatility which exploits more information in the data, unaffected by data arrival time and sampling frequency.

Using the assumption that asset prices are continuous semi-martingales, Malliavin and Mancino (2002, 2009) prove a general identity which relates the Fourier transform of the co-volatility function with the Fourier transform of the log price returns. We refer the reader to their papers for the full exposition and proofs. For our purposes, we have made use of their *integrated co-volatility estimator*, defined in Equation 1.

$$\hat{\Sigma}_{n,N}^{12} := \frac{1}{2N+1} \sum_{|s|<N} \sum_{i=0}^{n_1-1} \sum_{j=0}^{n_2-1} e^{(is(t_i^1-t_j^2))} \delta_{I_i^1}(p^1) \delta_{J_j^2}(p^2). \quad (1)$$

where $N$ is the number of Fourier coefficients, $n_k$ is the number of price changes for asset $k$ in the integrated window, $t_i^k$ is the time of each price change for asset $k$, $\delta_{I_i^k}(p^k)$ is the consecutive change in log-price of asset $k$ between time $t_i^k$ and $t_{i+1}^k$. To compute the required pairwise correlation, we compute the volatility for each asset, as well as the co-volatility for both, then use the simple relation in Equation 2.

$$\rho^{12} = \frac{\hat{\Sigma}_{n,N}^{12}}{\sqrt{\hat{\Sigma}_{n,N}^{11} \times \hat{\Sigma}_{n,N}^{22}}}. \tag{2}$$

We will apply this technique for finding *feature* correlations, i.e. considering *features* as *assets* in the above exposition. For the implementation, we used a combination of vectorisation and graphics processing unit (GPU) programming techniques to achieve efficient computation via parallelisation in MATLAB. Computation of complex exponential coefficients and multiplication with log price differences were performed on the GPU, after which the results in the *gpuArrays* were passed back to the CPU for final computation of co-volatility and correlation. Details of the implementation can be found in Wilcox et al. (2016).

## 4. High-speed feature clustering

A number of authors have promoted the use of spin glass models from statistical physics to capture the essential nature of complex financial systems (Blatt et al. (1996), Giada and Marsili (2002), Hendricks (2016)). In a recent paper, Hendricks et al. (2016b) provide an efficient computational solution for a technique initially proposed by Blatt et al. (1996, 1997) and Giada and Marsili (2001, 2002), where the super-paramagnetic ordering of a $q$-state Potts model is used for cluster identification. In a market Potts model, each stock can take on one of $q$ possible states, and each state can be represented by a cluster of similar stocks. Cluster membership is indicative of some commonality among the cluster members. Each stock has a component of its dynamics as a function of the state it is in and a component of its dynamics influenced by stock specific noise. In addition, there may be global couplings that influence all the stocks, i.e. the external field that represents a market mode.

We refer the reader to Giada and Marsili (2001), Hendricks et al. (2016b,a) for a comprehensive discussion of the technique and derivation of the pertinent log-likelihood fitness function. A key assumption in the derivation is that an object's price time series increments, $\bar{x}_i$, evolve under Noh (2000) model dynamics, whereby objects belonging to the same cluster share a common component, i.e.

$$\bar{x}_i = g_{s_i}\bar{\eta}_{s_i} + \sqrt{1 - g_{s_i}^2}\bar{\epsilon}_i, \tag{3}$$

where $\bar{\epsilon}_i$ is a vector describing the deviation of object $i$ from the features of cluster $s$, and $\bar{\eta}_{s_i}$ describes cluster-specific features. $g_s$ is a loading factor that encodes the similarity or difference between objects in cluster $s$.

Then, for a candidate cluster configuration of $n$ objects, $\mathcal{S} = \{s_1, ..., s_n\}$, the log-likelihood of $\mathcal{S}$ explaining the structure inherent in the data is given by

$$\mathcal{L}_c(\mathcal{S}) = \frac{1}{2} \sum_{s:n_s>1} \left( \log \frac{n_s}{c_s} + (n_s - 1) \log \frac{n_s^2 - n_s}{n_s^2 - c_s} \right), \tag{4}$$

where

$$C_{i,j} = \frac{\bar{x}_i \bar{x}_j}{\sqrt{\|\bar{x}_i^2\| \|\bar{x}_j^2\|}} \quad (5)$$

is the Pearson correlation between the $i^{th}$ and $j^{th}$ objects, representing the short-range distance-dependent ferromagnetic interaction term in the Potts analogy,

$$n_s = \sum_{i=1}^{N} \delta_{s_i,s} \quad (6)$$

is the number of objects in the $s^{th}$ cluster and

$$c_s = \sum_{i=1}^{N} \sum_{j=1}^{N} C_{i,j} \delta_{s_i,s} \delta_{s_j,s} \quad (7)$$

is the intra-cluster correlation.

Hendricks et al. (2016b) show that the likelihood function specified in Equation 4 can be used as an objective function in a high-speed, scalable parallel genetic algorithm (PGA), where candidate cluster configurations are evaluated and successively improved until a configuration best explains the inherent structure suggested by a correlation matrix. We will utilise their computational solution, since it provides the near-realtime efficiency required for our proposed online algorithm.

It is important to note that this approach, applied to stock features as objects, is consistent with the notion of financial markets as complex adaptive systems (Hendricks et al. (2016a), Wilcox and Gebbie (2014)). Essentially, if we consider the spin glass model analogy described in Blatt et al. (1996, 1997), we are finding a spin glass configuration (feature clustering) which permits a *metastable* system state given the spin-interaction term (feature correlations). This provides us with an unsupervised means to determine the optimal feature configuration *at the scale of interaction*, using a technique with is suited to the complex system dynamics.

## 5. Cluster configuration similarity and state discrimination

In Sections 3 and 4, we demonstrated a scheme to determine the feature cluster configuration from raw asynchronous data streaming from a market data feed, over an integrated window. Given this choice of state representation, what remains is to provide a feasible scheme to discriminate between states, such that the state space can be enumerated online.

This is a special case of a more general problem: measuring the distance between overlapping cluster configurations of a fixed set of objects. Consider two candidate cluster configurations of a fixed set of $n$ objects, $C_1 = \{s_1, s_2, ..., s_n\}$ and $C_2 = \{s'_1, s'_2, ..., s'_n\}$, where $s_k$ and $s'_k$ are the cluster indices to which the $k^{th}$ object belongs in each configuration. We would like to define a distance metric $d(C_1, C_2)$ which quantifies the configuration differences, while preserving the properties of symmetry, identity of indiscernables, non-negativity and sub-additivity.

Goldberg et al. (2010) considered this problem and proposed three candidate measures to quantify cluster configuration differences. We chose to implement a variant on their *best match* metric, which effectively counts the number of *moves* required to convert one configuration to the other. The measure is defined as,

$$d(C_1, C_2) = \frac{n}{n^2 - 1} \sum_{i=1}^{n} \min_{j} d(s_i, s'_j), \quad (8)$$

where $d(s, s') = 1 - \frac{|s \cap s'|}{|s \cup s'|}$ and $\frac{n}{n^2-1}$ is a normalisation constant, such that the maximum distance between two configurations is 1 (*single cluster* vs *all singletons*). For example, consider the following two candidate cluster configurations:

$$C_1 = \{1, 2, 3, 3, 4, 4, 4, 5\}$$
$$C_2 = \{1, 2, 2, 2, 3, 3, 4, 5\}.$$

Then, when considering cluster index 2, $d(s_2, s'_2) = 1 - \frac{|s_2 \cap s'_2|}{|s_2 \cup s'_2|} = 1 - \frac{1}{3} = 0.67$. Thus we have,

Table 1. Demonstration of *best match* metric for calculating distance between two overlapping cluster configurations

| Cluster index $k$ | $\|s_k \cap s'_k\|$ | $\|s_k \cup s'_k\|$ | $d(s_k, s'_k)$ |
|---|---|---|---|
| **1** | 1 | 1 | 0.00 |
| **2** | 1 | 3 | 0.67 |
| **3** | 0 | 4 | 1.00 |
| **4** | 1 | 3 | 0.67 |
| **5** | 1 | 1 | 0.00 |
| $d(C_1, C_2)$ | | | $\frac{2.34}{4.8} = \mathbf{0.49}$ |

Given a quantified distance between cluster configurations, we need to specify some *distance threshold* which encodes the idea that the configurations are sufficiently similar to be categorised as the same state. The specification of this *distance threshold* is somewhat ad-hoc and a source for subjective input, however coupled with a learning objective, we can iterate through multiple candidate thresholds to determine one which optimises the objective. In the use-case demonstrated in Section 8, we consider a simple *Q-learning* algorithm which aims to maximise wealth by deciding when to buy/sell shares. Starting with a given cash amount and stock inventory, we provide the learning agent with a fixed set of actions (buy/sell volumes) which it can perform in each period, based on the prevailing state. Each candidate distance threshold will provide the agent with a different lens to discriminate states, hence the state-action policy and terminal wealth of the agent will be impacted. We will choose a threshold which maximises terminal wealth.

The *best match* metric is easy to compute, intuitive and efficient, however the alternative metrics proposed by Goldberg et al. (2010) should be explored in further research, to assess the impact on state discrimination.

## 6. Reinforcement learning with online state discovery

Reinforcement learning (RL) is a technique for finding a calibrated policy in a controlled Markovian system with unknown dynamics, mapping system states to optimal or near-optimal decisions, given some objective. Learning can be model-free (Watkins (1989)) or model-based (Sutton (1990)), however the key principle is that feedbacks from interactions with the system can be used to provide insight for optimal planning decisions. We refer the reader to Kaelbling et al. (1996) for a comprehensive review of RL techniques. We will consider using the state space enumeration technique described in Sections 2 to 5 for an RL agent, enabling online planning decisions to be made with an adaptive state space.

The learning algorithm we consider can be seen as a particular implementation of the Dyna-Q architecture proposed by Sutton (1990), Sutton and Barto (1998), whereby feedbacks from the system are simultaneously used to improve the model of the system dynamics (state transitions), as well as the state-action policy for the learning objective. Both the feedbacks (or rewards) and expected state transitions are used to enumerate a so-called *Q-matrix* online, which contains the (current) discounted expected reward for each state-action pair, assuming the optimal policy is followed after the current time-step (Watkins (1989)). At the $t^{th}$ step in the learning algorithm, the agent:

- observes its current state $S_t \in \mathcal{S}$,
- selects and performs an action $A_t \in \mathcal{A}$,
- estimates the subsequent state $S_{t+1}$ as a result of performing action $A_t$ using the current model of state transitions,
- receives an immediate reward $r_t$ and
- uses a learning factor $\alpha_t$, which decreases gradually over time.

$Q$ is updated as follows:

$$Q(S_t, A_t) = Q(S_t, A_t) + \alpha_t [r_t + \gamma \max_b Q(S_{t+1}, b) - Q(S_t, A_t)], \tag{9}$$

where $\gamma$ is the *discount rate* controlling the importance of future rewards. In the usual discrete-actions, discrete-states specification, the *state space* $\mathcal{S}$ and permissible actions $\mathcal{A}$ are fixed *a priori* and remain fixed for the learning program. We now consider the case where the action set $\mathcal{A}$ remains fixed, but the state space $\mathcal{S}$ is dynamic, viz. discovered online as the agent interacts with the system. Each time a new state is *discovered* using the proposition in Section 5, the state space is increased. For the *Q-learning* update rule in Equation 9, the *next state* is determined using the prevailing empirical transition probability matrix, as

$$S_{t+1} = \arg\max_{S_j} Pr(S_t, S_j) \text{ for } S_j \in \mathcal{S}.$$

For the purposes of the investigation in Section 8, we assume that the agent's actions do not affect the system state evolution. This is an artefact of interacting with a *historical* data feed, where the consequences of an agent's actions cannot easily be incorporated. The overall proposition is, however, designed for a live trading agent submitting actual market orders, thus a *live* trading agent will affect the data feed it receives (absorbing limit orders through trades, affecting the LOB features), and thus the state space it perceives. We expect the efficacy demonstrated in Section 8 to translate to live trading.

## 7. Problem description and algorithm

### 7.1. *Wealth maximisation: Long-only*

To test the efficacy of the framework proposed in this paper, we construct a wealth maximising trading agent operating in high-frequency financial markets, able to buy and sell quantities of *one* given stock. The agent begins with a specified level of *cash* and *stock inventory*. At each trading opportunity (we assume regular 5-minute periods, however this can be generalised), the agent is able to *buy* certain quantities of the stock using available *cash*, or sell certain quantities of stock based on the level of *inventory*. We assume the agent is subject to a *long-only* (LO) constraint, i.e. the agent is not able to short-sell inventory, and is not allowed to use leverage. The *reward* is calculated as the portfolio PnL following the chosen action, i.e. the difference between the current portfolio value (inventory marked-to-market at current mid-price + cash) and the initial portfolio value. All price associated with all *buy* actions is the prevailing *best ask price*, and for all *sell* actions the prevailing *best bid price*. This ensures that *spread* is included as a transaction cost when trading. The agent uses the asynchronous tick-level data observed over the preceeding 5-minute window to compute feature correlations and the associated cluster configuration, as described in Sections 3 and 4, before determining the state, using the proposition in Section 5, and updating the *Q-matrix*.

## 7.2. *Algorithm*

Algorithm 1 describes the general implementation of the learning algorithm with online state discovery. We begin by initialising an empty *Q-matrix*, as no states have been discovered. After the first estimation period has passed (in this case 5 minutes), the raw feature data in the estimation period window is used to compute feature correlations, and then the feature cluster configuration. The correlation computation using a parallelised implementation of the Fourier estimator takes at most 0.05 seconds (Wilcox et al. (2016)), and the cluster configuration computation takes at most 0.8 seconds (Hendricks et al. (2016b)). This is the most expensive part of the algorithm, thus the achieved sub-second computation time is conducive for online application. We made use of fairly modest hardware with a gaming graphics processing unit (GPU) (Core i7-3820, 16GB DDR3 RAM, Nvidia Titan X 12GB GDDR5), thus there is room for further reductions in computation time through hardware scaling.

Following the cluster configuration computation, we compute the distance between the current feature configuration and the previously-identified configurations associated with prevailing states in the state space. If the distance between the current configuration and a previously-identified configuration is less than the specified threshold, we ascribe the associated state index to the current configuration. If the distance to all previously-identified configurations is larger than the threshold, a new state is created. The prevailing transition probability matrix is updated with the new state. We then use the $\epsilon$-*greedy* algorithm to choose an action based on the prevailing *Q-matrix*, recording the reward as the difference between the marked-to-market portfolio value and initial portfolio value. The transition probability matrix is then used to identify the next expected state, and the *Q-value* associated with the current state-action pair is updated using Equation 9.

---

**Algorithm 1** Unsupervised state detection and learning
---
Initialise *Q-matrix*
**while** Trading program not complete **do**
    Extract feature time-series (raw, asynchronous events) for integrated window
    Compute feature correlations using Fourier estimator
    Compute feature cluster configuration $\mathcal{C}$ using PGA
    Compare $\mathcal{C}$ with previously identified configurations $\mathcal{C}_i \in$ **state space**
    *Update state space*
    **if** *state space* $= \emptyset$ **then**
        Generate new state
    **else if** distance($\mathcal{C},\mathcal{C}_i$) $\leq$ *threshold* for any $\mathcal{C}_i \in$ **state space** **then**
        Assign state index of $\mathcal{C}_i$ to $\mathcal{C}$
    **else**
        Generate new state
    **end if**
    *Update state transition probability matrix*
    Update empirical prob of 1-step transition given all identified states
    **if** initialisation period complete **then**
        *Choose current optimal buy/sell action using Q-matrix*
        Record reward $R$ as difference between current MTM portfolio value
        and initial portfolio value
    **end if**
    *Update Q-matrix*
    Determine next state using current transition prob matrix
    Update Q-value associated with state-action pair, given recorded reward, next state
    and prevailing Q-value
**end while**

---

## 8. Data and Results

### 8.1. *Data*

The data for this study constituted tick-level trades and top-of-book quotes for one candidate stock on the Johannesburg Stock Exchange (JSE) from 1 October 2012 to 30 November 2012. We have chosen Anglo American (AGL) as an example of a liquid stock on the JSE with the typical trading volumes expected to support (automated) systematic strategies such as the proposal in this paper. This data was sourced from the Thomson Reuters Tick History (TRTH) database.

The raw data was stored in a MongoDB noSQL database, with appropriate indexes created for efficient retrieval and manipulation. The particular fields of interest for our study are: *Trade Price, Trade Volume, Level-1 Bid Price, Level-1 Bid Volume, Level-1 Ask Price, Level-1 Ask Volume*. Each of these features are represented by an unevenly-spaced time series in the dataset based on event occurrence. The stored, asynchronous event data is a close approximation to a stream of asynchronous events arriving from a live market data feed. We will apply our learning algorithm to this data in its most raw form, to avoid any subjective bias which may be introduced by data pre-processing techniques.

Table 2 shows the parameters which were used in the results which follow. A number of experiments with varying state discrimination thresholds were performed using a 1-week dataset (1 October 2012 to 5 October 2012). These initial results yielded a value of 0.05 which maximised average end-of-day PnL over this period. While we will explore the full sensitivity of PnL to this threshold in further research, here we will demonstrate the efficacy of the algorithm using a fixed threshold of 0.05.

Table 2. Parameters used for testing long-only wealth maximisation algorithm.

| Stock | AGL |
|---|---|
| Features | Trade Price/Volume, L1 Ask/Bid Price/Volume |
| Initial Cash | R100 000 |
| Initial Inventory | 800 shares |
| Buy Actions (proportion of Cash) | $\{0, 0.1, 0.2, ..., 0.9, 1.0\}$ |
| Sell Actions (proportion of Inventory) | $\{0.1, 0.2, 0.3, ..., 0.9, 1.0\}$ |
| State discrimination threshold | 0.05 |
| Probability of random action | 0.05 |
| Start time | 09:05 |
| End time | 16:30 |
| Estimation period / Trading frequency | 5 minutes |
| Initialisation period | 5 periods |

### 8.2. *Results*

Table 3 shows the summarised results from our analysis. We ran the algorithm for each day in our data set (01 October 2012 to 30 November 2012), constructing a distribution of end-of-day (16:30) PnL, expressed as a percentage of initial portfolio value. A positive percentage thus indicates that the algorithm decisions *created value* over the trading day. To test the efficacy of our algorithm, we compared its performance to a *random agent* and a *buy-and-hold agent*. The *random agent* chooses actions randomly at each decision point, i.e. makes no use of the learnt *Q-matrix*. This will allow us to test whether our algorithm is at least better than choosing actions at random. The *buy-and-hold agent* takes no trading actions, but rather holds the initial inventory for the duration of the trading day. This will allow us to test whether a learnt trading policy was better or worse than doing nothing at all. The table shows the minimum, lower quartile (LQ), mean, median, upper quartile (UQ), maximum and standard deviation of the end-of-day PnL distributions for the three agents.

Table 3. Summarised results from algorithm testing. The *LO Wealth Maximiser* agent is compared to a *Random* agent, where actions are chosen randomly at each trading opportunity, and a *Buy-and-Hold* agent, where no trading takes place and the initial inventory is held for the day. The algorithm begins at 09:05 and ends at 16:30 each trading day. These results summarise the distribution of end-of-day (16:30) PnL recorded for each day in the investigation period (01 Oct 2012 to 30 Nov 2012).

| **AGL** | \multicolumn{7}{c}{**PnL (as % of initial portfolio value)**} | | | | | | |
|---|---|---|---|---|---|---|---|
| *Model* | Min | LQ | **Mean** | **Median** | UQ | Max | Std Dev |
| *LO Wealth Maximiser* | -2.16 | -0.38 | **0.12** | **0.18** | 0.53 | 2.77 | 0.98 |
| *Random* | -2.62 | -1.10 | **-0.61** | **-0.48** | -0.24 | 1.30 | 0.87 |
| *Buy-and-Hold* | -3.87 | -0.83 | **-0.01** | **-0.22** | 0.65 | 3.19 | 1.39 |

Based on the results in Table 3, we see that the *LO wealth maximiser* agent generates a significantly better mean and median end-of-day PnL compared to both the *random agent* and *buy-* *and-hold agent* over this investigation period. In fact, the entire distribution for the *LO wealth maximiser* is more positively skewed, indicating that, in general, the agent generated a modest, but positive daily PnL. This is significant, as after only 5 periods of initialisation to refine the transition probability matrix, with no prior training, the agent is able to learn a useful policy fast enough to generate a positive PnL by the end of the trading day. While many more tests need to be run, varying the parameters in Table 2 and considering longer (and varied) investigation periods, these results indicate that the approach suggested in this chapter may be an effective framework for deploying purposeful trading agents which are able to identify exploitable structure in streaming market data feeds, and learn policies fast enough.

Figures 2 to 6 illustrate a typical run of the online algorithm, at various stages in the trading day, showing the identified states (top-left), prevailing transition probability matrix (top-right), portfolio PnL, stock mid-price, best bid and best ask levels (bottom-left) and current *Q-matrix* values (bottom-right). For the transition probability matrix and *Q-matrix* values, green indicates higher positive values (darker is higher), red/orange indicates lower values (darker is lower) and grey indicates an uninitialised state transition/state-action pair. In the current PnL plot, green dots indicate *buy* decisions and red dots indicate *sell* decisions, with the size of the dot being proportional to the quantity.

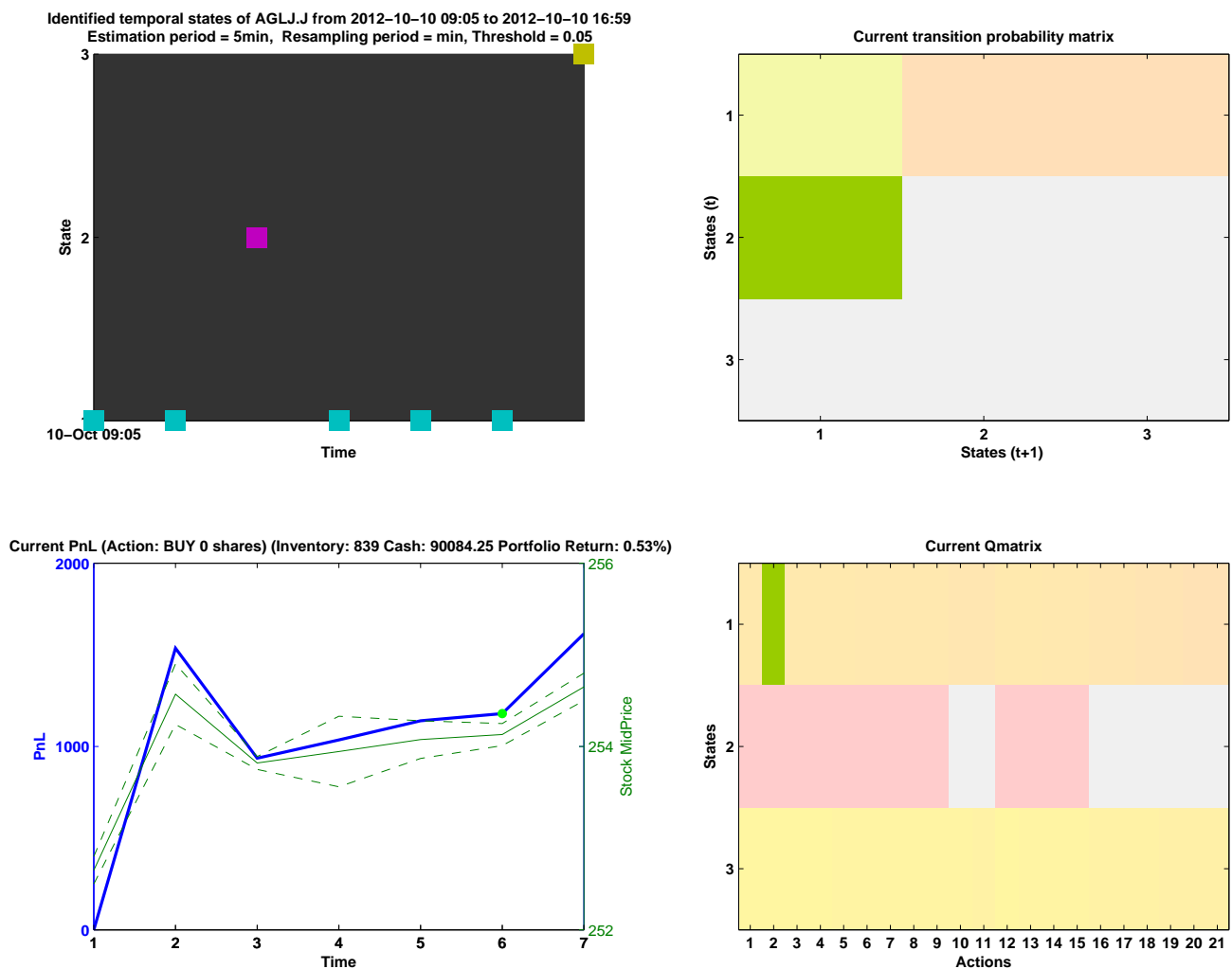

Figure 2. *Status at 09:35.* Demonstration of long-only wealth maximisation algorithm, starting with R100 000 cash and 800 AGL shares. The *top-left* plot shows the identified states since the start of the trading program (09:05), where blocks in the same row indicate the same state. The *top-right* plot illustrates the current empirical 1-step transition probability matrix, based on identified states. The *bottom-left* plot shows the stock mid-price, best ask and best bid in green (right Y-axis) and the running portfolio PnL in blue (left Y-axis). The portfolio PnL is determined by the difference between the current portfolio value (inventory marked-to-market at current mid-price + cash) and the initial portfolio value. Green dots indicate *buy* actions and red dots indicate *sell* actions, where the size of the dot is proportional to the quantity bought/sold. The *bottom-right* plot shows the current Q-matrix values, illustrating the expected cumulative discounted reward for each state-action pair.

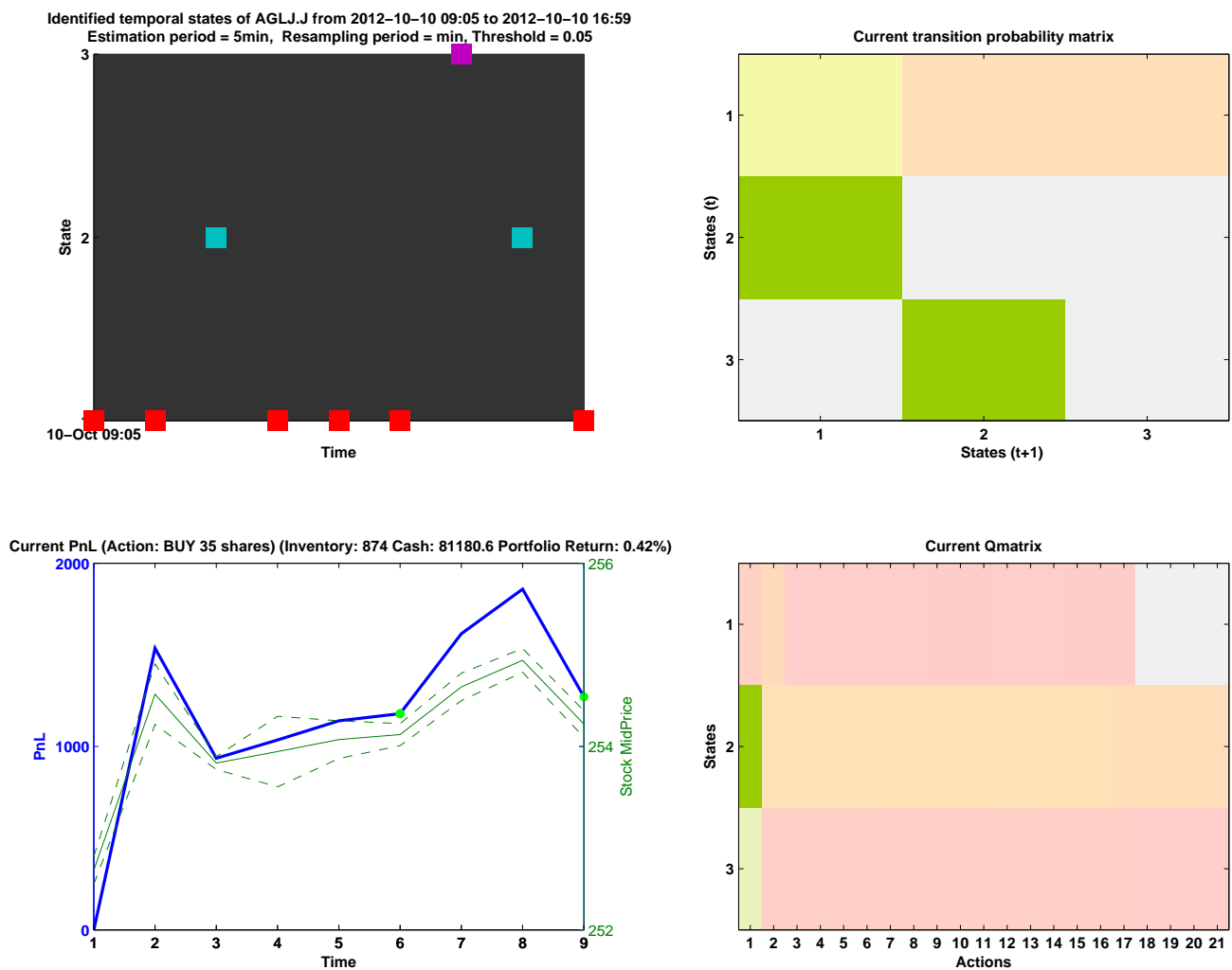

Figure 3. *Status at 09:45.* Demonstration of long-only wealth maximisation algorithm, starting with R100 000 cash and 800 AGL shares. The *top-left* plot shows the identified states since the start of the trading program (09:05), where blocks in the same row indicate the same state. The *top-right* plot illustrates the current empirical 1-step transition probability matrix, based on identified states. The *bottom-left* plot shows the stock mid-price, best ask and best bid in green (right Y-axis) and the running portfolio PnL in blue (left Y-axis). The portfolio PnL is determined by the difference between the current portfolio value (inventory marked-to-market at current mid-price + cash) and the initial portfolio value. Green dots indicate *buy* actions and red dots indicate *sell* actions, where the size of the dot is proportional to the quantity bought/sold. The *bottom-right* plot shows the current Q-matrix values, illustrating the expected cumulative discounted reward for each state-action pair.

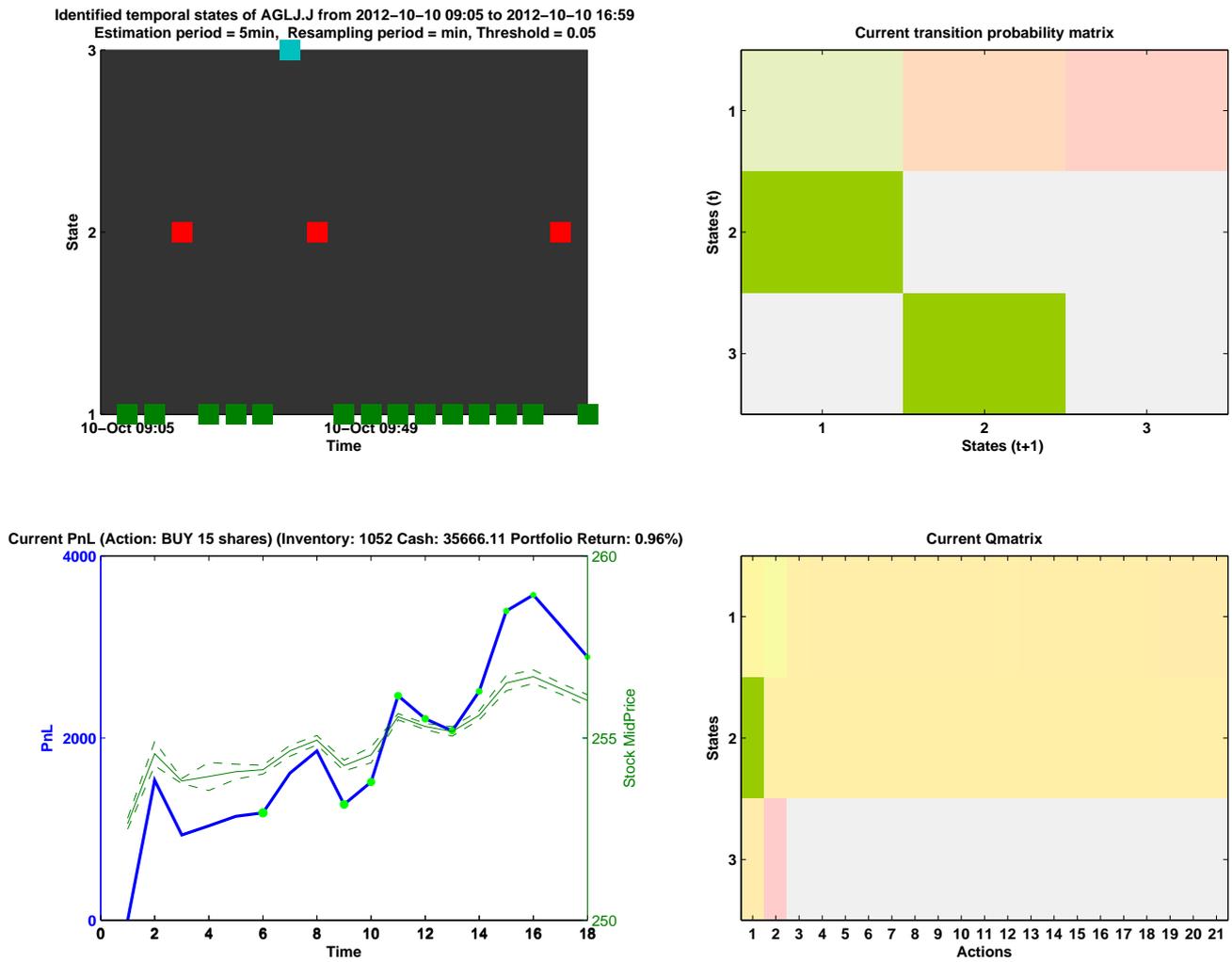

Figure 4. *Status at 10:30.* Demonstration of long-only wealth maximisation algorithm, starting with R100 000 cash and 800 AGL shares. The *top-left* plot shows the identified states since the start of the trading program (09:05), where blocks in the same row indicate the same state. The *top-right* plot illustrates the current empirical 1-step transition probability matrix, based on identified states. The *bottom-left* plot shows the stock mid-price, best ask and best bid in green (right Y-axis) and the running portfolio PnL in blue (left Y-axis). The portfolio PnL is determined by the difference between the current portfolio value (inventory marked-to-market at current mid-price + cash) and the initial portfolio value. Green dots indicate *buy* actions and red dots indicate *sell* actions, where the size of the dot is proportional to the quantity bought/sold. The *bottom-right* plot shows the current Q-matrix values, illustrating the expected cumulative discounted reward for each state-action pair.

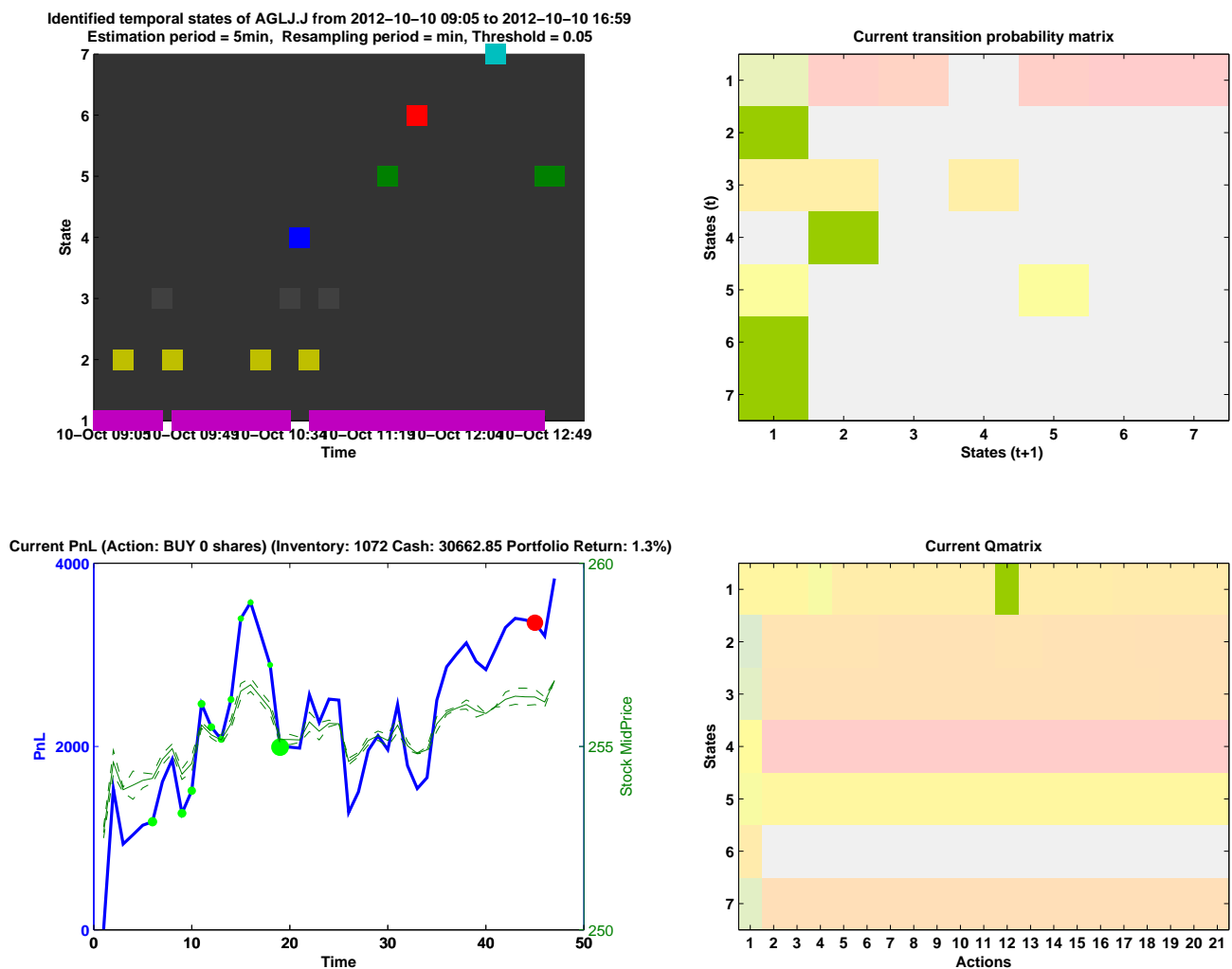

Figure 5. *Status at 12:45.* Demonstration of long-only wealth maximisation algorithm, starting with R100 000 cash and 800 AGL shares. The *top-left* plot shows the identified states since the start of the trading program (09:05), where blocks in the same row indicate the same state. The *top-right* plot illustrates the current empirical 1-step transition probability matrix, based on identified states. The *bottom-left* plot shows the stock mid-price, best ask and best bid in green (right Y-axis) and the running portfolio PnL in blue (left Y-axis). The portfolio PnL is determined by the difference between the current portfolio value (inventory marked-to-market at current mid-price + cash) and the initial portfolio value. Green dots indicate *buy* actions and red dots indicate *sell* actions, where the size of the dot is proportional to the quantity bought/sold. The *bottom-right* plot shows the current Q-matrix values, illustrating the expected cumulative discounted reward for each state-action pair.

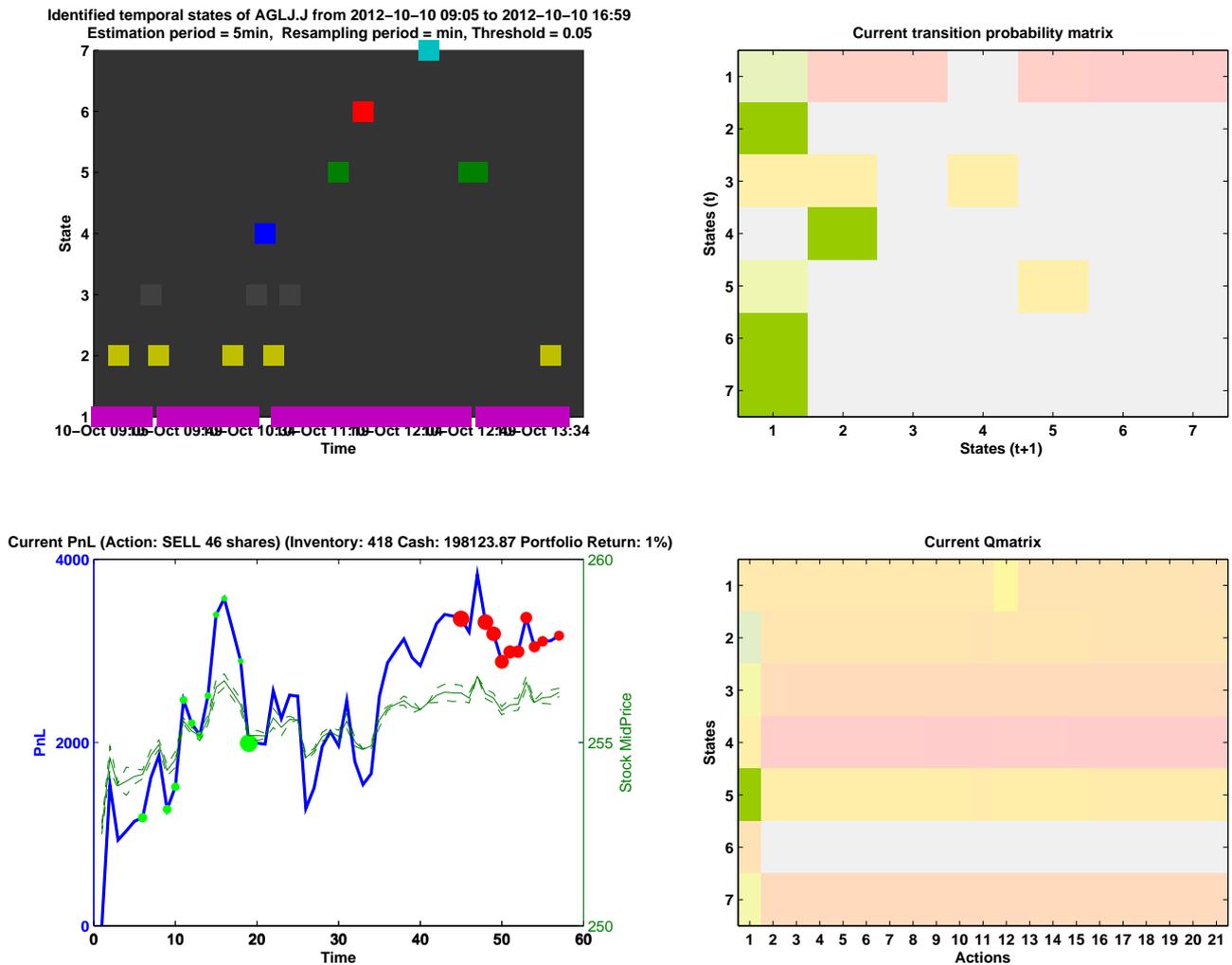

Figure 6. *Status at 13:30.* Demonstration of long-only wealth maximisation algorithm, starting with R100 000 cash and 800 AGL shares. The *top-left* plot shows the identified states since the start of the trading program (09:05), where blocks in the same row indicate the same state. The *top-right* plot illustrates the current empirical 1-step transition probability matrix, based on identified states. The *bottom-left* plot shows the stock mid-price, best ask and best bid in green (right Y-axis) and the running portfolio PnL in blue (left Y-axis). The portfolio PnL is determined by the difference between the current portfolio value (inventory marked-to-market at current mid-price + cash) and the initial portfolio value. Green dots indicate *buy* actions and red dots indicate *sell* actions, where the size of the dot is proportional to the quantity bought/sold. The *bottom-right* plot shows the current Q-matrix values, illustrating the expected cumulative discounted reward for each state-action pair.

## 9. Conclusion

In this paper, we demonstrated a scheme for online, unsupervised state discovery, detection and learning in high frequency markets, which is consistent with the complex adaptive system paradigm. By treating stock features as objects and applying the Potts model clustering approach, we are essentially finding a spin glass configuration (feature clustering) which permits a *metastable* system state given the spin-interaction term (feature correlations). Combined with a candidate state discrimination technique, this allows us to enumerate the state space online, permitting a purposeful agent to learn useful policies in an unknown domain without knowledge of human-preprocessed features and states.

The framework introduced here is conceptual and there are many areas for refinement, however we demonstrated that even with a simple choice of cluster distance metric and only top-of-book LOB features, a wealth-maximising agent can systematically outperform a random agent and buy-and-hold agent over the investigation period considered. In particular, after only 5 periods of initialisation to refine the transition probability matrix, with no prior training, the agent is able to learn a useful policy fast enough to generate a positive PnL by the end of the trading day. This is a promising result for the technique and bodes well for further research. We note that this technique is market agnostic and, in principle, could be applied to any system with a streaming, multi-featured data feed where the underlying system is a complex adaptive system.

We note that we have assumed that the agent's actions do not affect the system state evolution. This is an artefact of interacting with a *historical* data feed, where the consequences of an agent's actions cannot easily be incorporated. The overall proposition is, however, designed for a live trading agent submitting actual market orders, thus a *live* trading agent will affect the data feed it receives (absorbing limit orders through trades, affecting the LOB features), and thus the state space it perceives. We expect the efficacy of the algorithm to translate to live trading.

Prudent further investigations would include using alternative streaming features from the LOB (such as market depth), testing the stability of the algorithm over longer and more varied investigation periods, testing the sensitivity to the state discrimination threshold parameter, using *event-time* estimation windows / decision frequencies and alternative actions and reward functions.

## Acknowledgments

This work is based on the research supported in part by the National Research Foundation of South Africa (Grant number 89250). The conclusions herein are due to the author and the NRF accepts no liability in this regard. The author thanks Diane Wilcox and Tim Gebbie for their comments and suggestions, as well as the Fields Institute for Research in Mathematical Sciences and the University of Toronto for hosting him while much of the work for this paper was completed.

## References

Almgren, R., Chriss, N., 2000. Optimal execution of portfolio transactions. Journal of Risk **3**, 5–39.

Bengio, Y., Courville, A., Vincent, P., . Representation learning: A review and new perspectives. IEEE Transactions on Pattern Analysis and Machine Intelligence .

Bertsimas, D., Lo, A., 1998. Optimal control of execution costs. Journal of Financial Markets **1**, 1–50.

Blatt, M., Wiseman, S., Domany, E., 1996. Superparamagnetic clustering of data. Phys. Rev. Lett. **76**, 3251–3254.

Blatt, M., Wiseman, S., Domany, E., 1997. Data clustering using a model granular magnet. Neural Computation **9**, 1805–1842.

Boulanger-Lewandowski, N., Bengio, Y., Vincent, P., 2012. Modeling temporal dependencies in high-dimensional sequences: Application to polyphonic music generation and transcription. Proceedings from the International Conference on Machine Learning (ICML) .

Cartea, A., Jaimungal, S., Penalva, J., 2015. Algorithmic and high-frequency trading. Cambridge University Press, Cambridge, UK.

Ciresan, D., Meier, U., Masci, J., Gambardella, L., Schmidhuber, J., 2011. Flexible, high performance convolutional neural networks for image classification. Proceedings of the Twenty-Second International Joint Conference on Artificial Intelligence (IJCAI) .

Ciresan, D., Meier, U., Schmidhuber, J., 2012. Multi-column deep neural networks for image classification. Working paper URL: http://arxiv.org/abs/1202.2745.

Dahl, G., Yu, D., Deng, L., Acero, A., 2012. Context-dependent pre-trained deep neural networks for large vocabulary speech recognition. IEEE Transactions on Audio, Speech, and Language Processing **20**, 33–42.

Forsyth, P., 2011. A hamilton-jacobi-bellman approach to optimal trade execution. Applied Numerical Mathematics **61**, 241–265.

Frei, C., Westray, N., 2015. Optimal execution of a VWAP order: A stochastic control approach. Mathematical Finance **25**, 612–639.

Giada, L., Marsili, M., 2001. Data clustering and noise undressing of correlation matrices. Phys. Rev. E **63**.

Giada, L., Marsili, M., 2002. Algorithms of maximum likelihood data clustering with applications. Physica A **315**, 650–664.

Glorot, X., Bordes, A., Bengio, Y., 2011. Domain adaptation for large-scale sentiment classification: A deep learning approach. Proceedings from the International Conference on Machine Learning (ICML) .

Goldberg, M., Hayvanovych, M., Magdon-Ismail, M., 2010. Measuring similarity between sets of overlapping clusters. Proceedings from the International Conference on Social Computing (SocialCom) , 303–308.

Griffin, J., Oomen, R., 2011. Covariance measurement in the presence of non-synchronous trading and market microstructure noise. Journal of Econometrics **160**, 58–68.

Hendricks, D., 2016. An online adaptive learning algorithm for optimal trade execution in high-frequency markets. Ph.D. thesis. University of the Witwatersrand. URL: http://hdl.handle.net/10539/21710.

Hendricks, D., Gebbie, T., Wilcox, D., 2016a. Detecting intraday financial market states using temporal clustering. Quantitative Finance 16, 1657–1678. URL: http://dx.doi.org/10.1080/14697688.2016.1171378.

Hendricks, D., Gebbie, T., Wilcox, D., 2016b. High-speed detection of emergent market clustering via an unsupervised parallel genetic algorithm. South African Journal of Science 112. URL: http://dx.doi.org/10.17159/sajs.2016/20140340.

Hendricks, D., Wilcox, D., 2014. A reinforcement learning extension to the Almgren-Chriss framework for optimal trade execution. Proceedings from IEEE Conference on Computational Intelligence for Financial Economics and Engineering URL: http://dx.doi.org/10.1109/CIFEr.2014.6924109.

Hinton, G., 2007. Learning multiple layers of representation. Trends in Cognitive Sciences **11**, 428–434.

Hinton, G., Deng, L., Dahl, G., Mohamed, A., Jaitly, N., Senior, A., Vanhoucke, V., Nguyen, P., Sainath, T., Kingsbury, B., 2012. Deep neural networks for acoustic modeling in speech recognition. IEEE Signal Processing Magazine **29**, 82–97.

Kaelbling, L., Littman, M., Moore, A., 1996. Reinforcement learning: A survey. Journal of Artificial Intelligence Research **4**, 237–285.

Krizhevsky, A., Sutskever, I., Hinton, G., 2012. Imagenet classification with deep convolutional neural networks. Proceedings from Neural Information Processing Systems (NIPS) conference .

Lee, H., Grosse, R., Ranganath, R., Ng, A., 2009. Convolutional deep belief networks for scalable unsupervised learning of hierarchical representations. Proceedings from the International Conference on Machine Learning (ICML) .

Malliavin, P., Mancino, M., 2002. Fourier series method for measurement of multivariate volatilities. Finance and Stochastics **6**, 49–61.

Malliavin, P., Mancino, M., 2009. A Fourier transform method for nonparametric estimation of multivariate volatility. Annals of Statistics **37**, 1983–2010.

Mnih, V., Kavukcuoglu, K., Silver, D., Graves, A., Antonoglou, I., Wierstra, D., Riedmiller, M., 2013. Playing Atari with deep reinforcement learning. Neural Information Processing (NIPS) deep learning workshop .

Nevmyvaka, Y., 2004. Normative approach to market microstructure analysis. Ph.D. thesis. Carnegie Mellon University.

Nevmyvaka, Y., Feng, Y., Kearns, M., 2006. Reinforcement learning for optimal trade execution. Proceedings of the 23rd international conference on machine learning .

Noh, J., 2000. A model for correlations in stock markets. Physical Review E **61**.
Sutton, R., 1990. Integrated architectures for learning, planning, and reacting based on approximating dynamic programming. Proceedings of the Seventh International Conference on Machine Learning , 216–224.
Sutton, R., Barto, A., 1998. Reinforcement learning. MIT Press, Cambridge, MA.
Watkins, C., 1989. Learning from delayed rewards. Ph.D. thesis. Cambridge University.
Wilcox, D., Gebbie, T., 2014. Hierarchical causality in financial economics. Working paper URL: http://ssrn.com/abstract=2544327.
Wilcox, D., Hendricks, D., Gebbie, T., 2016. Fourier methods for correlation estimation. Working paper .